# Voltage-induced high-speed DW motion in a synthetic antiferromagnet

Lulu Chen[1], Maokang Shen[1#], Yingying Peng[1], Xinyu Liu[1], Wei Luo[1], Xiaofei Yang[1], Yue Zhang[1*]

1.*School of Optical and Electronic Information, Huazhong University of Science and Technology, Wuhan, 430074, PR China*

*Corresponding author: yue-zhang@hust.edu.cn (Yue Zhang)

#Maokang Shen has the same contribution to Lulu Chen

**Abstract**

Voltage-induced motion of a magnetic domain wall (DW) has potential in developing novel devices with ultralow dissipation. However, the speed for the voltage-induced DW motion (VIDWM) in a single ferromagnetic layer is usually very low. In this work, we proposed VIDWM with high speed in a synthetic antiferromaget (SAF). The velocity for the coupled DWs in the SAF is significantly higher than its counterpart in a single ferromagnetic layer. Strong interlayer antiferromagnetic exchange coupling plays a critical role for the high DW velocity since it inhibits the tilting of DW plane with strong Dzyaloshinskii-Moriya interaction. On the other hand, the Walker breakdown of DW motion is also inhibited due to the stabilization of moment orientation under a strong interlayer antiferromagnetic coupling. In theory, the voltage-induced gradient of magnetic anisotropy is proved to be equal to an effective magnetic field that drives DW.

Motion of magnetic domain walls (DWs) exhibits significant potential in development of wide novel magneto electronic devices, such as racetrack memory, magnetic logic devices, and spin memristors for mimicking synapse in artificial neural network [1-4]. Voltage-induced DW motion (VIDWM) has outstanding advantage for low dissipation, and it attracts wide research attention [5-9]. Nevertheless, the speed of voltage-induced DW motion (VIDWM) in a single ferromagnetic nanotrack is not quite high, and complicated fabrication technology seems necessary to accelerate the VIDWM [10].

A high DW velocity is expectable for VIDWM in a synthetic antiferromagnet (SAF), in which a high velocity (750 m/s or higher) of current-induced DW motion (CIDWM) has been reported [11, 12]. Either current or voltage only results in an initial DW motion, and the ultrahigh velocity is mainly attributed to strong antiferromagnetic exchange coupling between two ferromagnetic layers and the interfacial Dzyaloshinskii-Moriya Interaction (DMI) between a heavy metal layer and a ferromagnetic (FM) layer [11]. Very recently, voltage-induced motion of coupled skyrmions in an SAF at an ultrahigh velocity was proposed [13]. Nevertheless, the device based on DW motion seems easier in technique. In this work, VIDWM in the SAF is studied numerically and clear increase in DW velocity and inhibition of Walker breakdown has been observed and explained in theory.

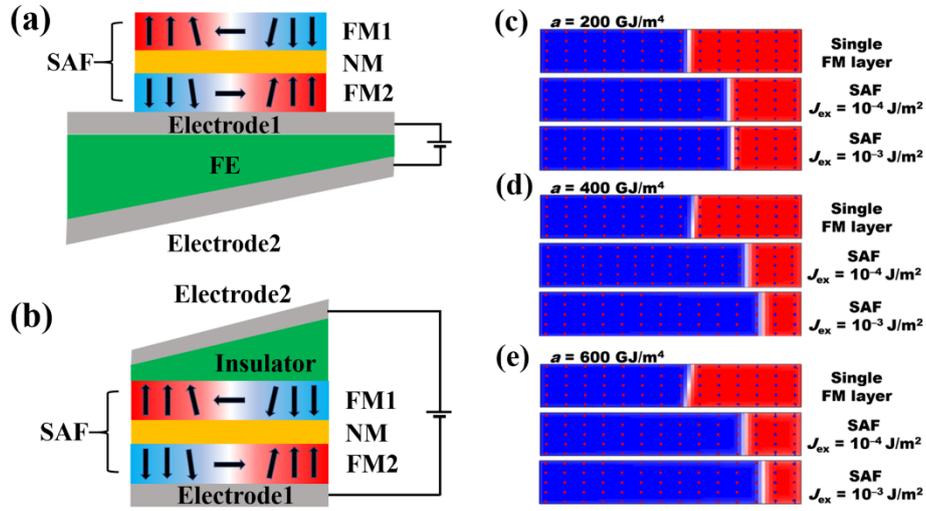

Fig. 1. Schematic of voltage-induced DW motion based on multiferroic behavior (a) and manipulation of charge state at metal/insulator interface (b); (c) - (e). DW motion under different voltage-induced gradient of magnetic anisotropy energy in a single FM layer and in the SAFs with different interlayer exchange coupling.

The schematic of VIDWM in the SAF is shown in Fig. 1. In principle, the VIDWM is based on manipulation of magnetic anisotropy constant under an external voltage. In application, two routes can be exploited. One is based on the multiferroic behavior with the combination of converse piezoelectric effect and magnetostriction effect (Fig. 1 (a)) [5]. The other is originated from the variation of charge state at the metal/insulator interface under an external electric field (Fig. 1 (b)) [6-9]. In either case, a wedge-shaped piezoelectric substrate or insulator layer is necessary for inducing an electric field strength ($E$) increasing from one end to the other under a voltage. This sloped $E$ gives rise to spatial variation of magnetic anisotropy energy, and the DW in the ferromagnetic layer move towards the end with a lower anisotropy energy [5, 14]. In the SAF, the DWs in the upper and lower layer move together when the interlayer exchange coupling is strong enough.

The Object-Oriented Micromagnetic Framework (OOMMF) software with the code of interfacial DMI was exploited to do the micromagnetic simulation [15]. The model is a track of SAF multilayer with perpendicular magnetic anisotropy (PMA). The parameters are as follows. The length and width of the track are 600 nm and 100 nm, respectively. The thicknesses of the lower layer, the upper layer, and the interlayer are all 0.6 nm. The cell dimension is 2 nm × 1 nm × 0.6 nm. The saturation magnetization ($M_S$) is $5.8 \times 10^5$ A/m; the damping coefficient $\alpha$ is 0.03. The DMI constant ($D$) varies between 0 mJ/m$^2$ and 3 mJ/m$^2$. The interlayer Ruderman–Kittel–Kasuya–Yosida (RKKY) exchange coupling parameters ($J_{ex}$) between two FM layers is between 0 J/m$^2$ and $-1 \times 10^{-3}$ J/m$^2$. The magnetic anisotropy constant ($K$) varies as a linear function of $x$, $K = ax + b$. Here $a$ is the gradient of anisotropy constant, between 0 GJ/m$^4$ and 600 GJ/m$^4$. The DW motion in a single FM layer with the same parameters was also simulated for comparison.

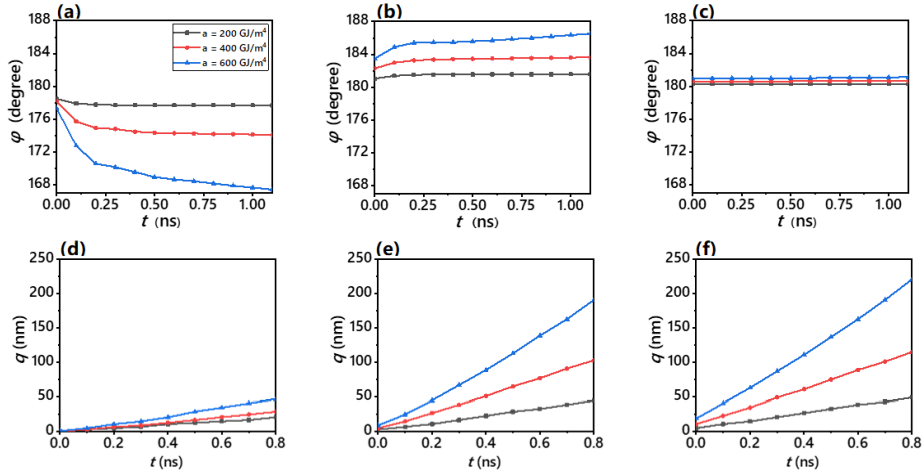

Fig. 2. Azimuthal angle and central position of DW as a function of time under different gradient of magnetic anisotropy energy in a single FM layer ((a) and (d)) and in the SAF with $J_{ex} = -1 \times 10^{-4}$ J/m$^2$ ((b) and (e)) and $J_{ex} = -1 \times 10^{-3}$ J/m$^2$ ((c) and (f)).

Initially, the DW is generated in the middle of track. After the gradient of $K$ is applied on the track, the DW starts to move towards the end with lower $K$. In the single FM layer, the DW velocity is smaller than 100 m/s and the DW plane tilts when the gradient of $K$ reaches 600 GJ/m$^4$ due to strong

DMI [16-19]. However, in the SAF, the DWs move clearly faster and the DW tilting is inhibited by the interlayer RKKY interaction (Fig. 1(b)-(d)). The DW velocity increases with increasing gradient of $K$ and the interlayer RKKY coupling.

It is also noticed the DW moves at an almost constant velocity when $a$ is as small as 200 GJ/m$^4$, but it moves at an ever-increasing speed when $a$ becomes 600 GJ/m$^4$ (Fig. 2). This accelerated motion does not appear for the current-induced DW motion and is special for the voltage-induced DW motion [14]. In general, the DW velocity is strictly related to the temporal azimuthal angle of the moment in the central of DW, and it becomes stable when the azimuthal angle reaches a constant value. In a single FM layer, the azimuthal angle keeps decreasing in the process of DW motion, especially for the large gradient of $K$ (Fig. 2(a)). In the SAF with strong interlayer exchange coupling, changing of azimuthal angle with time is depressed significantly, and the azimuthal angle for the moment in the central of DW becomes close to 180 degrees. This indicates strong interlayer exchange coupling in the SAF contributes to the stabilization of DW motion and the Néel-type DW structure even though the gradient of $K$ is as large as 600 GJ/m$^4$.

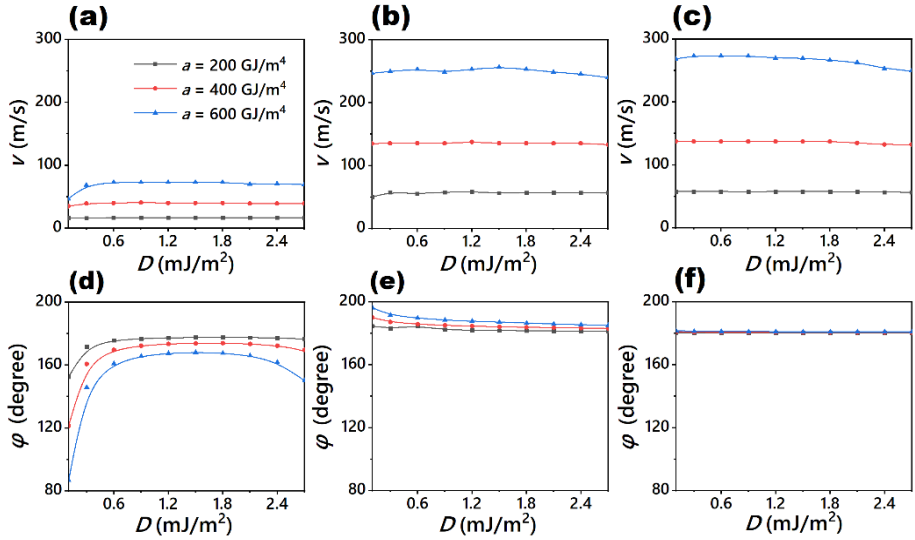

**Fig. 3. Average DW velocity and azimuthal angle as a function of DMI constant under different gradient of magnetic anisotropy energy in a single FM layer ((a) and (d)) and in the SAF with $J_{ex} = -1 \times 10^{-4}$ J/m$^2$ ((b) and (e)) and $J_{ex} = -1 \times 10^{-3}$ J/m$^2$ ((c) and (f)).**

In addition to interlayer-exchange coupling, DMI is another important factor influencing the DW motion. In a single FM layer, average DW velocity and $D$ satisfies nonmonotonous relationship, and maximum DW velocity appears at certain $D$ (Fig. 3). This nonmonotonous behavior also works for the variation of azimuthal angle with $D$, but it is clearly depressed under strong interlayer exchange coupling. In application, DMI is a double-edge sword for DW motion: A moderate DMI offers a torque that contributes to the driving force of DW motion [20], but a strong DMI tilts the DW plane that reduces the DW motion [21]. Strong interlayer exchange coupling inhibits the DW tilting and enables the DW to keep its high velocity under a strong DMI.

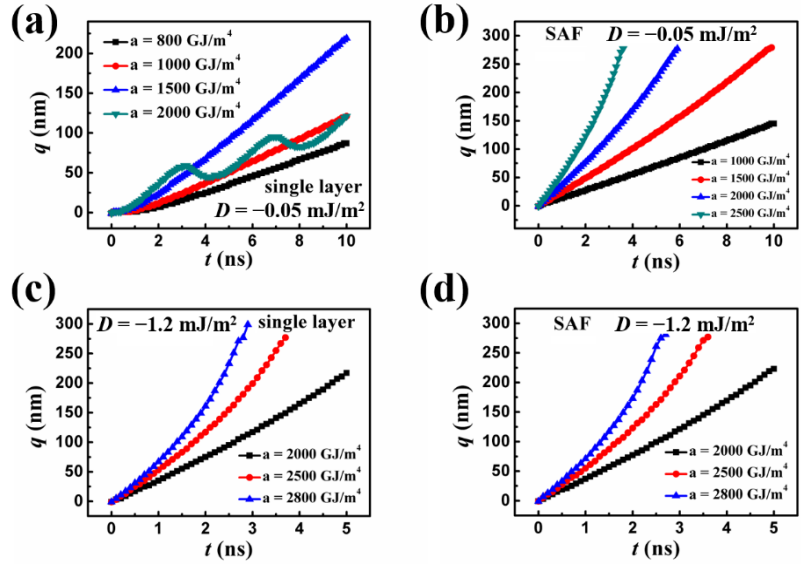

**Fig. 4. Time dependence of central position of DW in a single layer for (a) $D = -0.05$ mJ/m$^2$ and (c) $D = -1.2$ mJ/m$^2$ and an SAF for (b) $D = -0.05$ mJ/m$^2$ and (d) $D = -1.2$ mJ/m$^2$ ($J_{ex} = -10^{-3}$ J/m$^2$).**

In VIDWM, a large gradient of magnetic anisotropy energy may introduce Walker breakdown [5]. The stabilization of DW structure under strong interlayer exchange coupling hints the possibility for inhibiting Walker breakdown. To prove this, we have done the simulation for DW motion in a single layer and SAFs with different DMI under large gradient of magnetic anisotropy constant ($a$) (Fig. 4). In a single layer with $D = -0.05$ mJ/m$^2$, Walker breakdown occurs when $a$ is 2000 GJ/m$^4$ (Fig. 4(a)), but it is not observed in an SAF with $J_{ex} = -10^{-3}$ J/m$^2$ even when $a$ is as high as 2500 GJ/m$^4$ (Fig. 4(b)). In the single layer and the SAF with stronger DMI ($D = -1.2$ mJ/m$^2$), the Walker breakdown is depressed. This is consistent with the theoretical prediction about depression of Walker breakdown by enhancing DMI [22].

In theory, DW motion driven by the gradient of $K$ can be depicted by the collective coordinate method. For simplification, the DW width is approximately taken as a constant, and the gradient of $K$ is converted to an effective magnetic field that drives DW. The collective coordinates, including the position, the azimuthal angle, and the polar angle of the moment in the central of DW, are labeled as $q_{L(U)}$, $\varphi_{L(U)}$, and $\theta_{L(U)}$, respectively. Here the subscript L and U denote the lower and upper layer, respectively. In a spherical coordinate system, the unit vector for the direction of magnetization is $\vec{m}_{L(U)} = (\sin\theta_{L(U)}\cos\varphi_{L(U)}, \sin\theta_{L(U)}\sin\varphi_{L(U)}, \cos\theta_{L(U)})$, and the ansatz is expressed as [11]:

$$\theta_L = 2\arctan\{\exp[(x-q(t))/\Delta]\}, \quad \varphi_L = \varphi_L(t), \tag{1}$$

and

$$\theta_U = 2\arctan\{\exp[(x-q(t))/\Delta]\} + \pi, \quad \varphi_U = \varphi_U(t). \tag{2}$$

Here $\Delta$ is the DW width that is approximately written as:

$$\Delta = \sqrt{A/(b-\frac{1}{2}\mu_0 M_S^2)}. \tag{3}$$

The DW dynamics is depicted quantitatively through the Gilbert equation:

$$\frac{d\vec{m}_i}{dt} = -\gamma \vec{m}_i \times (\vec{H}_{eff})_i + \alpha \vec{m}_i \times \frac{d\vec{m}_i}{dt}, \quad (i = L, U) \tag{4}$$

Here $\vec{H}_{eff}$ is the effective magnetic field that is related to the free energy density $w$ as:

$$(\vec{H}_{eff})_i = -\frac{1}{\mu_0 M_s}\frac{\delta w_i}{\delta \vec{m}_i}, \tag{5}$$

and

$$w_i = \frac{A}{\Delta^2}\sin^2\theta_i + (b-\frac{1}{2}\mu_0 M_S^2)\sin^2\theta_i + \frac{1}{2}\mu_0 N_x M_S^2 \sin^2\theta_i \cos^2\varphi_i \\ + \frac{D\sin\theta_i \cos\varphi_i}{\Delta} - \frac{J_{ex}}{t_s}[\sin\theta_L \sin\theta_U \cos(\varphi_L - \varphi_U) + \cos\theta_L \cos\theta_U] - \mu_0 M_S (H_z)_i \cos\theta_i \tag{6}$$

The $(H_z)_i$ represents the effective magnetic field for the gradient of magnetic anisotropy. The absolute value of $H_z$ of the two layers are the same [14]:

$$H_z = \frac{1}{2M_S}\frac{\partial \sigma}{\partial x} = \frac{2}{M_S}\frac{d\sqrt{AK_{eff}}}{dx} = \frac{a\sqrt{A}}{M_S}(ax+b)^{-\frac{1}{2}}. \tag{7}$$

However, the signs of $H_z$ for the two layers are opposite due to the opposite transition of magnetization in the two layers when two coupled DWs move in the same direction. Eqs. 5, 6, and 7 were put into Eq. 4 and the group of Thiele equations were derived after the integration for the

whole track:

$$\frac{2\alpha}{\Delta}\dot{q} + \dot{\varphi}_L - \dot{\varphi}_U = -\frac{\gamma a}{\Delta M_S}\int\sqrt{\frac{A}{ax+b}}\sin^2\theta dx; \qquad (8)$$

$$\frac{1}{\Delta}\dot{q} - \alpha\dot{\varphi}_L = -\frac{\gamma\pi D\sin\varphi_L}{2\Delta M_S} - \frac{\gamma\mu_0 N_x M_S \sin 2\varphi_L}{2} + \frac{\gamma J_{ex}}{M_S t_s}\sin(\varphi_U - \varphi_L); \qquad (9)$$

$$-\frac{1}{\Delta}\dot{q} - \alpha\dot{\varphi}_U = -\frac{\gamma\pi D\sin\varphi_U}{2\Delta M_S} + \frac{\gamma\mu_0 N_x M_S \sin 2\varphi_U}{2} + \frac{\gamma J_{ex}}{M_S t_s}\sin(\varphi_L - \varphi_U). \qquad (10)$$

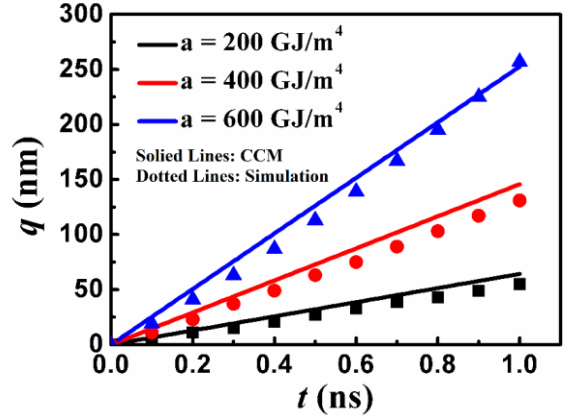

**Fig. 5. Central position as a function of time under different time (The solid lines are the numerical solution of CCM and the dotted lines are the result of micromagnetic simulation.)**

After a numerical approximation of the integration in Eq. 8, the group of Thiele equations were resolved numerically with the initial condition $q$ = 0 nm, $\varphi_L$ = 180º, and $\varphi_U$ = 0º. The numerical solution is close to that obtained from the simulation (Fig. 4). The small deviation may come from the approximation for the constant DW width. On the other hand, the track in the simulation has finite length, and the DW motion slows down when it approaches the track end [14].

In summary, a high-speed VIDWM in an SAF is predicted based on an electric-field-induced gradient of magnetic anisotropy. The velocity for the coupled DWs in the SAF is clearly higher than its counterpart in a single FM layer. Strong interlayer RKKY exchange coupling plays a critical role for the high-speed VIDWM since it inhibits the tilting of DW plane with strong DMI and the Walker breakdown. In theory, the gradient of magnetic anisotropy is proved to be equal to an effective magnetic field that drives DW. In application, the VIDWM may be realized based on multiferroic behaviors [23, 24] or voltage-induced variation of charge state at the metal/insulator interface [25].

**Acknowledgements**

This work was supported by the National Natural Science Foundation of China [grant number 11574096] and Huazhong University of Science and Technology (No. 2017KFYXJJ037).

**References**

1. Stuart S. P. Parkin, M. Hayashi, and L. Thomas, Science **320**, 190 (2008).
2. D. A. Allwood, G. Xiong, C. C. Faulkner, D. Atkinson, D. Petit, and R. P. Cowburn, Science **309**, 1688 (2005).
3. N. Locatelli, V. Cros, and J. Grollier, Nat. Mater. **13**, 11 (2014).
4. M. Sharad, C. Augustine, G. Panagopoulos, and K. Roy, Trans. Nanotech. **11**, 4 (2012).
5. K. Yamada, S. Murayama, and Y. Nakatani, Appl. Phys. Lett. **108**, 202405 (2016).
6. A. J. Schellekens, A. van den Brink, J. H. Franken, H. J. M. Swagten, and B. Koopmans, Nat. Commun. **3**, 847 (2012).
7. W. Lin, N. Vernier, G. Agnus, K. Garcia, B. Ocker, W. Zhao, E. E. Fullerton, and D. Ravelosona, Nat. Commun. **7**, 13532 (2016).
8. F. Ando, H. Kakizakai, T. Koyama, K. Yamada, M. Kawaguchi, S. Kim, K.-J. Kim, T. Moriyama, D. Chiba, and T. Ono, Appl. Phys. Lett. **109**, 022401 (2016).
9. D. Chiba, M. Kawaguchi, S. Fukami, N. Ishiwata, K. Shimamura, K. Kobayashi, and T. Ono, Nat. Commun. **3**, 888 (2012).
10. F. N. Tan, W. L. Gan, C. C. I. Ang, G. D. H. Wong, H. X. Liu, F. Poh, and W. S. Lew, Sci. Rep. **9**, 7369 (2019).
11. S. H. Yang, K. S. Ryu, and S. Parkin, Nat. Nanotech. **10**, 221 (2015).
12. Z. Yu, Y.Zhang, Z. Zhang, M. Cheng, Z. Lu,X. Yang, J. Shi, and R. Xiong, Nanotechnology **29**, 175404 (2018).
13. C. C. I. Ang ,W. Gan, and W. S. Lew, New J. Phys. **21**, 043006 (2019).
14. Y. Zhang, S. Luo, X.Yang, and C, Yang, Sci. Rep. **7**, 2047 (2017).
15. S. Rohart and A. Thiaville, Phys. Rev. B **88**, 184422 (2013).
16. O. Boulle, S. Rohart, L. D. Buda-Prejbeanu, E. Jué, I. M. Miron, S. Pizzini, J. Vogel, G. Gaudin, and A. Thiaville, Phys. Rev. Lett. **111**, 217203 (2013).
17. E. Martinez, S. Emori, N. Perez, L. Torres, and G. S. D. Beach, J. Appl. Phys. **115**, 213909 (2014).
18. K.-S. Ryu, L. Thomas, S.-H. Yang, and S. S. P. Parkin, Appl. Phys. Express **5**, 093006 (2012).
19. M. Shen, Y. Zhang, W. Luo, L. You, and X. Yang, J. Magn. Magn. Mater. **485**, 69 (2019).
20. S. H. Yang and S. Parkin, J. Phys. Condens. Matter **29**, 303001 (2017).
21. M. Shen, Y. Zhang, L. You, and X. Yang, Appl. Phys. Lett. **113**, 152401 (2018).
22. A. Thiaville, S. Rohart, É. JUÉ, V. Cros, and A. Fert, EPL **100**, 57002 (2012).
23. M. Liu, O. Obi, Z. Cai, J. Lou, G. Yang, K. S. Ziemer, and N. X. Sun, J. Appl. Phys. **107**, 073916 (2010).
24. M. Liu, O. Obi, J. Lou, Y. Chen, Z. Cai, S. Stoute, M. Espanol, M. Lew, X. Situ, K. S. Ziemer, V. G. Harris, and N. X. Sun, Adv. Funct. Mater. **19**, 1826 (2009).

25. W. G. Wang, M. Li, S. Hagemen, and C. L. Chien, Nat. Mater. **11**, 64, (2012).